\documentclass[english,10pt,final,twocolumn]{elsarticle}
\usepackage[T1]{fontenc}
\usepackage[latin9]{inputenc}
\usepackage{pdfpages}
\usepackage{amstext}
\usepackage{graphicx}
\usepackage{gensymb}
\usepackage{placeins}
\makeatletter

\newcommand{\lyxmathsym}[1]{\ifmmode\begingroup\def\b@ld{bold}
  \text{\ifx\math@version\b@ld\bfseries\fi#1}\endgroup\else#1\fi}
\usepackage{braket}
\usepackage{mathrsfs}
\usepackage{slashed}
\usepackage{bm}
\usepackage{datetime}
\usepackage{siunitx}

\DeclareSIUnit[number-unit-product = {}]\clight{c}
\DeclareSIUnit\eVperc{\eV\per\clight}
\DeclareSIUnit\GeVpercs{\giga\eV\squared\per\clight\squared}
\DeclareSIUnit\MeVpercs{\mega\eV\per\clight\squared}
\journal{Physics Letters B}
\biboptions{numbers,sort&compress}
\usepackage[left=1.6cm,right=1.04cm,top=1.65cm,bottom=1.65cm,columnsep=25pt]{geometry}
\usepackage{lineno}
\usepackage{hyperref}  
\hypersetup{breaklinks = true, colorlinks = true, allcolors = magenta}
\usepackage{setspace}
\usepackage{graphicx}
\usepackage{wrapfig}

\makeatother

\usepackage{babel}
\begin{document}

\begin{frontmatter}{}

\title{
Quasi-elastic polarization-transfer measurements \\ on the deuteron in anti-parallel kinematics
}

\author[TAU]{S.~Paul\corref{cor2}}

\ead{paulsebouh@mail.tau.ac.il}

\author[TAU]{D.~Izraeli}
\author[JSI]{T.~Brecelj}
\author[TAU]{I.~Yaron}

\author[Mainz]{P.~Achenbach}
\author[Mainz]{H.~Arenh\"ovel}
\author[TAU]{A.~Ashkenazi}
\author[JSI]{J.~Beri\v{c}i\v{c}}
\author[Mainz]{R.~B\"ohm}
\author[zagreb]{D.~Bosnar}
\author[Rutgers]{E.~Cline}
\author[TAU]{E.O.~Cohen}
\author[JSI]{L.~Debenjak}
\author[Mainz]{M.O.~Distler}
\author[Mainz]{A.~Esser}
\author[zagreb]{I.~Fri\v{s}\v{c}i\'{c}\fnref{mit}}
\author[Rutgers]{R.~Gilman}
\author[Konstanz]{M.~Heilig}
\author[Mainz]{S.~Kegel}
\author[Mainz]{P.~Klag}
\author[Mainz]{Y.~Kohl}
\author[JSI]{T.~Kolar}
\author[TAU,nrc]{I.~Korover}
\author[TAU]{J.~Lichtenstadt}
\author[TAU,soreq]{I.~Mardor}
\author[Mainz]{H.~Merkel}
\author[Mainz]{D.G.~Middleton}
\author[JSI,Mainz,UL]{M.~Mihovilovi\v{c} }
\author[Mainz]{J.~M\"uller}
\author[Mainz]{U.~M\"uller}
\author[TAU]{M.~Olivenboim}
\author[TAU]{E.~Piasetzky}
\author[Mainz]{J.~Pochodzalla}
\author[huji]{G.~Ron}
\author[Mainz]{B.S.~Schlimme}
\author[Mainz]{M.~Schoth}
\author[Mainz]{F.~Schulz}
\author[Mainz]{C.~Sfienti}
\author[UL,JSI]{S.~\v{S}irca}
\author[JSI]{S.~\v{S}tajner }
\author[USK]{S.~Strauch}
\author[Mainz]{M.~Thiel}
\author[Mainz]{A.~Tyukin}
\author[Mainz]{A.~Weber}

\author{\\\textbf{(A1 Collaboration)}}

\cortext[cor2]{Corresponding author}

\fntext[mit]{Present address: MIT-LNS, Cambridge, MA 02139, USA.}

\address[TAU]{School of Physics and Astronomy, Tel Aviv University, Tel Aviv 69978,
Israel.}

\address[JSI]{Jo\v{z}ef Stefan Institute, 1000 Ljubljana, Slovenia.}

\address[Mainz]{Institut f\"ur Kernphysik, Johannes Gutenberg-Universit\"at, 55099
Mainz, Germany.}

\address[zagreb]{Department of Physics, Faculty of Science, University of Zagreb, HR-10000
Zagreb, Croatia.}

\address[Rutgers]{Rutgers, The State University of New Jersey, Piscataway, NJ 08855,
USA.}

\address[Konstanz]{Universit\"at Konstanz, Fachbereich Physik, Universit\"atsstra{\ss}e 10, 78464 Konstanz, Germany.}

\address[nrc]{Department of Physics, NRCN, P.O. Box 9001, Beer-Sheva 84190, Israel.}

\address[soreq]{Soreq NRC, Yavne 81800, Israel.}

\address[huji]{Racah Institute of Physics, Hebrew University of Jerusalem, Jerusalem
91904, Israel.}

\address[UL]{Faculty of Mathematics and Physics, University of Ljubljana, 1000
Ljubljana, Slovenia.}

\address[USK]{University of South Carolina, Columbia, SC 29208, USA.}

\begin{abstract}
We present measurements of the polarization-transfer components in the $^2$H$(\vec e,e'\vec p)$ reaction, covering a previously unexplored kinematic region with large \textit{positive} (anti-parallel) missing momentum, $p_{\rm miss}$, up to 220 MeV$\!/\!c$, and $Q^2=0.65$ $({\rm GeV}\!/\!c)^2$.  
These measurements, performed at the Mainz Microtron (MAMI), were motivated by theoretical calculations which predict small final-state interaction (FSI) effects in these kinematics, making them favorable for searching for medium modifications of bound nucleons in nuclei.
We find in this kinematic region that the measured polarization-transfer components $P_x$ and $P_z$ and their ratio agree with the theoretical calculations, which use free-proton form factors.  Using this, we establish upper limits on possible medium effects that modify the bound proton's form factor ratio $G_E/G_M$ at the level of a few percent.
We also compare the measured polarization-transfer components and their ratio for $^2$H to those of a free (moving) proton. 
We find that the universal behavior of $^2$H, $^4$He and $^{12}$C in the double ratio $\frac{(P_x/P_z)^A}{(P_x/P_z)^{^1\!\rm H}}$ is maintained in the positive missing-momentum region.

\end{abstract}
\date{\today}

\end{frontmatter}{}
\newcommand{\mov}{M}
\section{Introduction}
Polarization transfer from a polarized electron to a proton in elastic scattering has
become a recognized method to measure the ratio of the proton's elastic electromagnetic form factors,
$G_E/G_M$ \cite{Jones:1999rz,Gayou:2001qd,Punjabi:2005wq,Milbrath:1997de,Barkhuff:1999xc,Pospischil:2001pp,PhysRevC.64.038202,MACLACHLAN2006261,PhysRevC.74.035201}.   Assuming the one-photon exchange approximation, the ratio of the transverse ($P_x$) 
to longitudinal ($P_z$) polarization-transfer components is proportional to $G_E/G_M$ \cite{Akh74}.
This provides a direct measurement of the form factor (FF) ratio and eliminates many systematic uncertainties \cite{Perdrisat}.  

Measuring the ratio of the same components of the polarization transfer to a \textit{bound} proton in \textit{quasi-free}
kinematics on nuclei, which is sensitive to the electromagnetic FF ratio,
has been suggested as a method to study differences between free and bound protons \cite{Milbrath:1997de,Barkhuff:1999xc}.  
As such it can be used as a tool to identify
medium modifications in the bound proton's internal structure, reflected in the FFs and thereby in the
polarization transfer.  Deviations between measured polarization ratios in quasi-free and elastic scattering can be interpreted only by comparing the measurements with realistic calculations of nuclear effects.  
This makes the deuteron an appropriate target for such measurements, since its nucleons can be highly virtual and it is a well understood nucleus from a theoretical standpoint \cite{Arenhovel}.

Quasi-free polarization-transfer experiments have been carried out on $^2$H and $^{12}$C at the
Mainz Microtron (MAMI) \cite{deep2012PLB,deepCompPLB,ceepLet}, as well as on $^2$H, $^4$He and $^{16}$O at Jefferson Lab (JLab) \cite{jlabDeep,Strauch,Paolone,Malov_O16}, in search of medium modification in the proton's internal structure. 
These include measurements on far off-shell nucleons, characterized by high missing momentum, which is equivalent (neglecting final-state interactions (FSI)) to the struck protons having high initial momentum.  
For $^2$H in parallel kinematics (negative $p_{\rm miss}$), it was shown that the deviations in $P_x/P_z$ from that of elastic $\vec ep$ scattering can be explained by
nuclear effects without the necessity of introducing modified FFs \cite{deep2012PLB,deepCompPLB}.  
The individual components $P_x$ and $P_z$ were also measured.  The calculated values of these components deviated from the measurements, suggesting that the FSI effects are not fully reproduced \cite{deepCompPLB}.  

In this work, we present the measurements of $P_x$ and $P_z$, and their ratio, in a previously unexplored kinematic region, where theoretical calculations predict that the FSI effects would be small compared to those in the kinematic regions of previous measurements (e.g. \cite{deep2012PLB,deepCompPLB}).  This makes this kinematic region useful for searching for genuine medium modifications, which would be obscured by large FSI effects in less favorable kinematical regions.  

\begin{figure}[th]
\begin{center}
\includegraphics[width=\columnwidth]{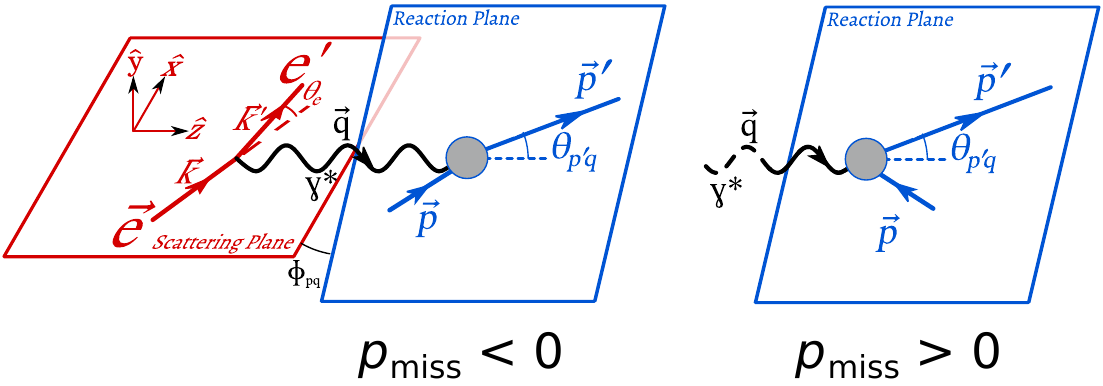}
\caption{
Kinematics with $p_{\rm miss} < 0$ (left) contrasted with kinematics with  $p_{\rm miss} > 0$ (right).  In the kinematics on the left ($p_{\rm miss}<0$), the proton is initially moving in nearly the same direction as the virtual photon, while in the kinematics on the right ($p_{\rm miss}<0$), the proton is initially moving in the opposite direction of the virtual photon, and then changes direction by up to 180 degrees.  
}
\label{fig:kinematics}
\end{center}
\end{figure}

In the $A(\vec e,e'\vec p)$ reaction, a polarized electron with a known initial four-momentum $k$ scatters off a bound proton with initial four-momentum $p$, through exchange of a virtual photon with four-momentum $q$.  The final momenta of the knock-out proton,  $\vec p^{\,\prime}$, and the scattered electron, $\vec k^{\,\prime}$, are measured.   We define the ``missing momentum'' of the reaction to be $\vec{p}_{\rm miss}=\vec{q}-\vec p^{\,\prime}$, where $\vec q=\vec{k}-\vec k ^{\,\prime}$ is the momentum transfer.   In the absence of FSI, the initial proton momentum is given by the missing momentum, $\vec{p}=-\vec{p}_{\rm miss}$.  We refer to the missing momentum as being positive (negative) if $\vec{p}_{\rm miss}\cdot\vec q$ is positive (negative).  (See Fig.~\ref{fig:kinematics}.)

Previous measurements of $^2$H$(\vec e,e'\vec p)$  were obtained at $Q^2 = 0.40$ $({\rm GeV}\!/\!c)^2$ with positive $p_{\rm miss}$ reaching  
175 MeV$\!/\!c$ \cite{deep2012PLB,deepCompPLB}.   
The new data presented in this work obtained at $Q^2 = 0.65$ $({\rm GeV}\!/\!c)^2$ have improved statistical precision and extend the probed positive $p_{\rm miss}$ range up to 220 MeV$\!/\!c$.
This new kinematical region allows a more meaningful test for universality by comparing deviations in the ratio $P_x/P_z$ between $^2$H and a free proton to those of $^{12}$C and $^4$He
 in $p_{\rm miss}>0$ kinematics, complementing the earlier tests at $p_{\rm miss}<0$ in \cite{deep2012PLB,deepCompPLB,ceepLet}.

\section{Experimental Setup}
\label{sec:setup}
The experiment was performed at MAMI using the A1 beamline and spectrometers \cite{a1aparatus}.  For these measurements, a 690 MeV polarized CW electron beam with a $\sim$9 $\mu$A current was used. The average beam polarization was $\sim$80\%, measured twice daily with a M\o ller polarimeter \cite{Wagner,Bartsch}.  These measurements were cross-checked with a Mott polarimeter \cite{Tioukine}, and the two measurements were consistent with one another, up to a renormalization factor.  In the analysis, we considered the time-dependence of the beam polarization, which is described by a fit of the M\o ller  measurements using two linear segments (see the supplementary material).  

 The beam helicity was flipped at a rate of 1 Hz. The target consisted of an oblong shaped cell (50 mm long, 11.5 mm diameter) filled with liquid deuterium. Two high-resolution, small solid-angle spectrometers with momentum acceptances of 20-25\% were used to detect the scattered electrons and knocked out protons in coincidence. The proton spectrometer was equipped with a polarimeter located close to the focal plane (FPP)  with a 7 cm thick carbon analyzer \cite{a1aparatus,Pospischil:2000pu}. The spin-dependent scattering of the polarized proton by the carbon analyzer allows the determination of $P_x$ and $P_z$ at the reaction point in the target \cite{Pospischil:2000pu} by correcting for the spin precession  in the spectrometer magnetic field.

In the analysis, cuts were applied to identify coincident electrons and protons that originate from the deuterium target, and to ensure good reconstruction of tracks in the spectrometer and FPP.  Only events where the proton scatters by more than 8$\degree$ in the FPP were selected (to remove Coulomb-scattering events).

We present measurements at two kinematic settings, labeled E and F, given in Table \ref{tab:kinematics}.  These complement previous measurements \cite{deep2012PLB,deepCompPLB}, the kinematic settings of which are listed in the supplementary materials.  This work primarily focuses on the data taken at setting E, with large positive $p_{\rm miss}$ and $Q^2=0.65$ $({\rm GeV}\!/\!c)^2$, where theoretical calculations (see Section \ref{sec:calculations}) predict smaller FSI effects than in previous settings.  

\begin{table}[h]
\caption{
The kinematic settings of the $^2$H($\vec{e},e'\vec p$) data presented in this work. The angles and momenta represent the center values for the two spectrometer setups. Also shown is the range of $\theta_{pq}$, the angle between $\vec{q}$ and $\vec p = -\vec{p}_{\rm miss}$. For the kinematics of the earlier $^2$H data sets, see the supplementary materials.}
\begin{center}
\begin{tabular}{lllll}
\hline
\multicolumn{2}{c}{} & Kinematic Setup &  \\ \cline{3-4}
 & &  E & F\\

\hline
$E_{\rm beam}$ & [MeV] & 690 & 690\\
$Q^2$ & [$({\rm GeV}\!/\!c)^2$] & 0.65 & 0.40 \\
$p_{\rm miss}$ & [MeV$\!/\!c$] & 60 to 220 & $-$70 to 70\\
$p_e$ & [MeV$\!/\!c$] & 464 & 474  \\
$\theta_e$ & [$\degree$] & 90.9  & 67.1 \\
$p_p$ & [MeV$\!/\!c$] & 656  & 668 \\
$\theta_{p}$ & [MeV$\!/\!c$] & $-$33.6 & $-$40.8 \\
$\theta_{pq}$ & [$\degree$] & 130 to 180 & 0 to 180 \\
\multicolumn{2}{l}{\small{\# of events after cuts}} & 138 k & 595 k  \\
\hline
\end{tabular}
\end{center}
\label{tab:kinematics}
\end{table}%

We also took data at $Q^2=0.40$ $({\rm GeV}\!/\!c)^2$ and $p_{\rm miss}\sim 0$ (setting F) during the same run-period.  The larger cross-section in setting F allowed us to obtain large statistics, and to compare the measurements with those at a similar kinematic setting (A) in Refs.~\cite{deep2012PLB,deepCompPLB} as a cross-check.
\FloatBarrier

\section{Measured polarization-transfer}
\begin{figure}[ht!]
\begin{center}
\includegraphics[width=\columnwidth]{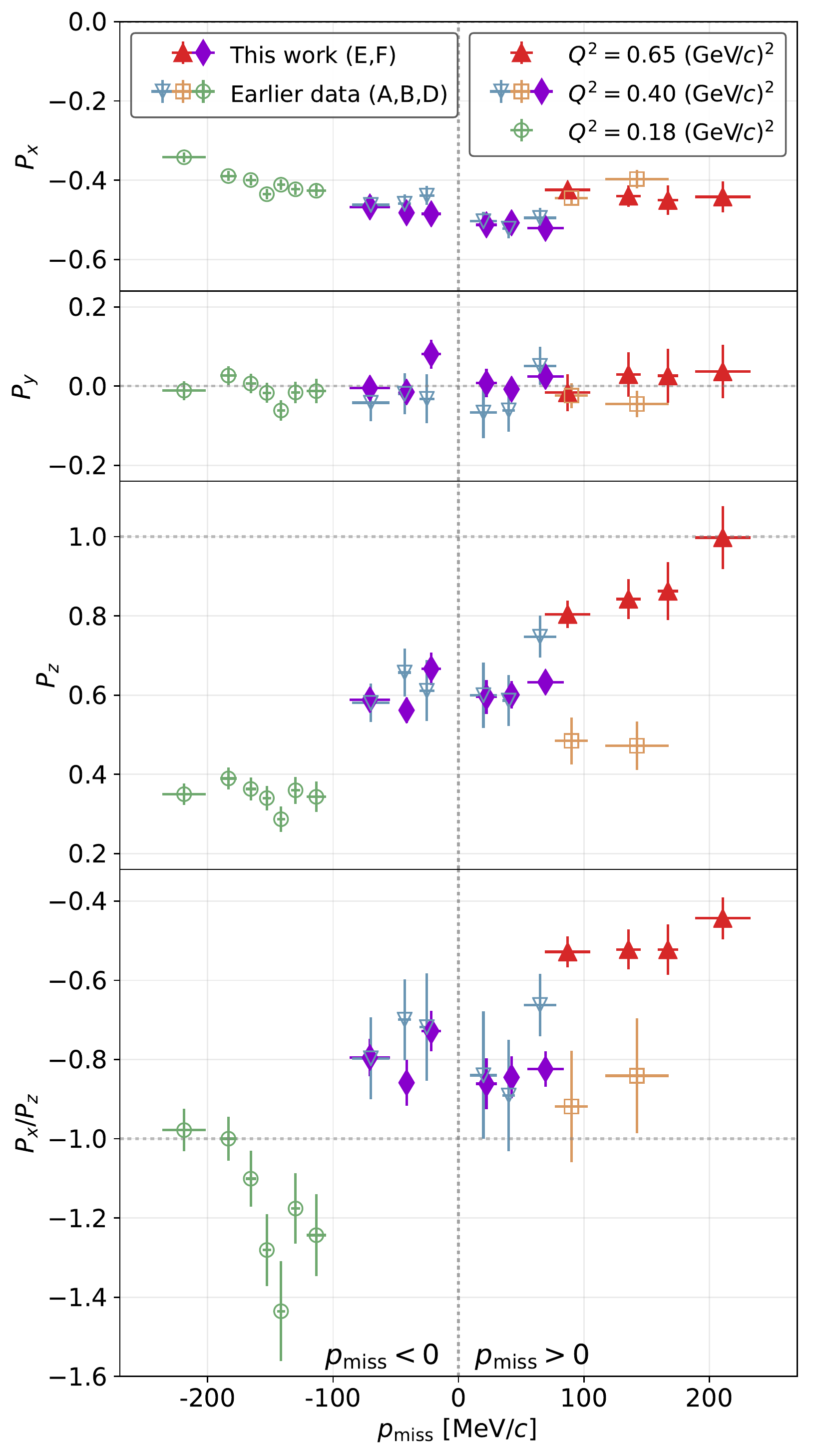}
\caption{
The measured polarization components $P_x$ (top), $P_y$ (second panel) and $P_z$ (third panel), and the ratio $P_x/P_z$ (bottom) measured in this work (red and purple, filled), and the earlier $^2$H data sets \cite{deep2012PLB,deepCompPLB} (open symbols).  The vertical error bars here (and in the other figures)  are statistical only.  The horizontal bars represent the standard deviation of the $p_{\rm miss}$ in the bin.  
}
\label{fig:raw_r_vs_pmiss}
\end{center}
\end{figure}

The polarization-transfer components, $P_x$ and $P_z$, and their ratio $P_x/P_z$ measured in setups E and F, along with the earlier $^2$H data from MAMI \cite{deep2012PLB,deepCompPLB}, are shown in Fig.~\ref{fig:raw_r_vs_pmiss}.  The new data in setting E ($Q^2=0.65$ $({\rm GeV}\!/\!c)^2$) extend the range of our positive $p_{\rm miss}$ data, matching the coverage of the earlier data in the negative $p_{\rm miss}$ region (setting D, at $Q^2=0.18$ $({\rm GeV}\!/\!c)^2$).     

The errors shown in Fig.~\ref{fig:raw_r_vs_pmiss} are statistical only.  The systematic error in these measurements is due to a few sources. The
largest contribution to the uncertainty in the polarization components
$P_x$ and $P_z$ is the beam polarization. It was measured periodically during
the data taking by Moller and Mott polarimeters with an estimated
accuracy of 2\% (see the supplementary material). The carbon analyzing power
in this kinematic region is known to about 1\% \cite{Pospischil:2000pu,AprileGiboni:1984pb,Mcnaughton:1986ks}. However, the 
ratio $P_x/P_z$ is independent of the analyzing power and beam polarization.
The uncertainty in the spin-precession evaluation is dominated by the
reaction vertex reconstruction, and was evaluated to contribute 0.8\%.
The reconstruction of the momentum and scattering angle of the detected
particles contribute 0.4\%. The uncertainty due to the beam energy is 0.5\%. 
Contributions due to detector alignment are negligible. Helicity-independent
uncertainties, such as target density, detector acceptance and efficiency, etc.,
largely cancel out due to the frequent flips of the beam helicities.
Software cuts applied in the analysis were studied and contribute 1.5\%.
The total estimated systematic uncertainties are 3\%, and 2\% for the
polarization components and their ratio respectively.

We find that $P_x$ shows mild variation between different kinematics, but $P_z$ varies greatly between data sets, approaching 1 at large $p_{\rm miss}$ for setting E.  
The $P_y$ component, which vanishes in elastic scattering \cite{Akh74}, is nearly zero for each of our data sets, indicating that this component is not strongly affected by nuclear effects.

The polarization-transfer components for setting F ($p_{\rm miss}\sim0$, $Q^2=0.4$ $({\rm GeV}\!/\!c)^2$) agree within the errors with the earlier measurements which were obtained in a similar kinematic\footnote{These data sets have the same $Q^2$ and cover the same range in $p_{\rm miss}$, but had different beam energies: 600 (690) MeV$\!/\!c$ for setting A (F).} (setting A), providing further support that the normalization due to beam polarizations is also within the error.   However, settings E and B (which overlap in positive $p_{\rm miss}$ but have different $Q^2$: 0.65 and 0.40 $({\rm GeV}\!/\!c)^2$ respectively) are very different from one another in $P_z$.  

A more coherent picture is obtained by dividing the measured polarization components by the expected ones for a free-proton, and by examining them as functions of the struck-proton virtuality, as discussed in Sec.~\ref{sec:calculations} below.

\section{Comparison of measurements to elastic $\vec ep$ scattering and $^2$H$(\vec e,e'\vec p)$ calculations}
\label{sec:calculations}

\begin{figure}[th]
\includegraphics[width=.97\columnwidth]{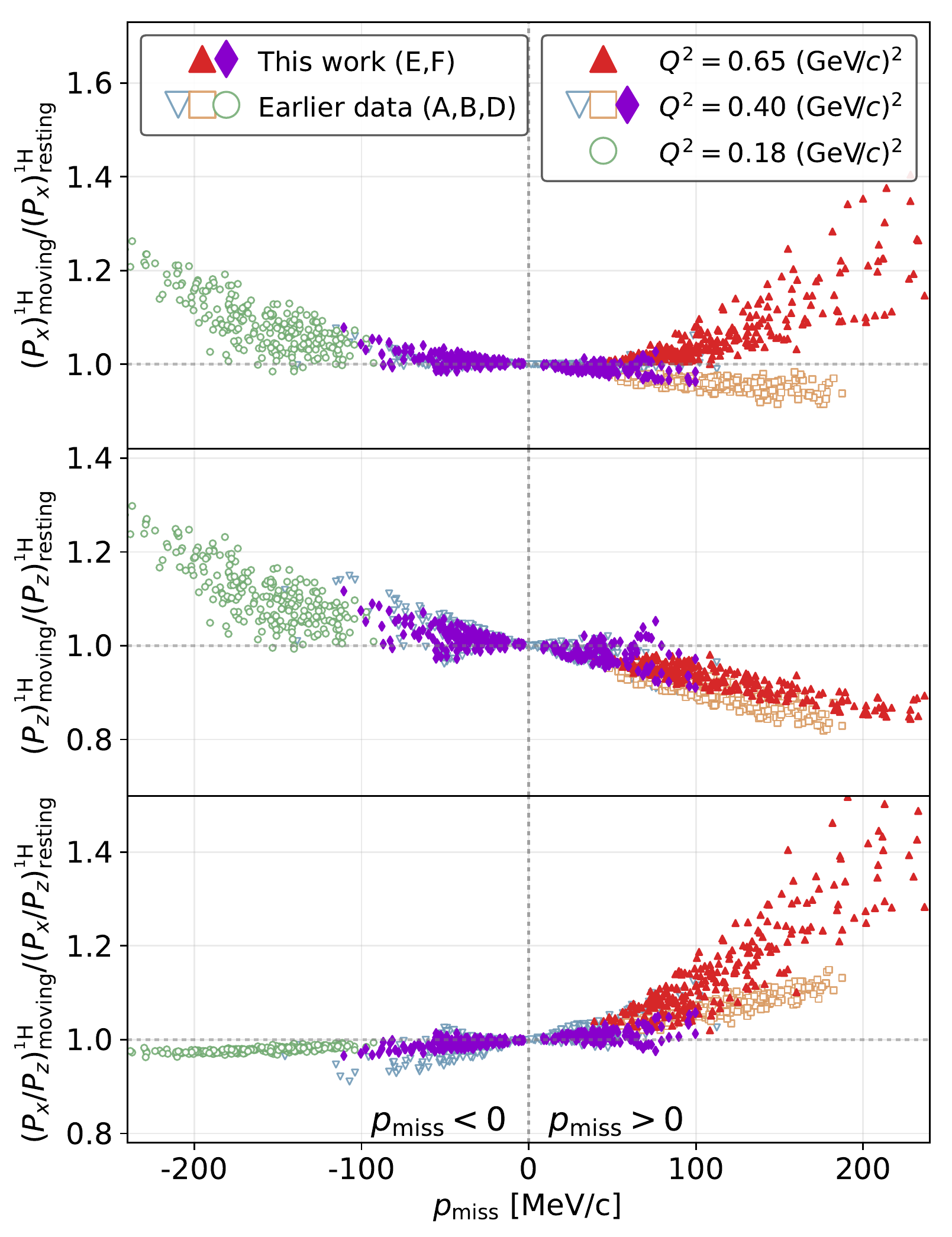}
\caption{Ratios of $P_x$, $P_z$ and $P_x/P_z$ for moving free protons \cite{ArenhovelMoving,movingProtonPLB} to those for free resting protons \cite{Akh74}, using the kinematics derived from measured events in $^2\textrm{H}(\vec{e},e'\vec p)\textrm{n}$ reactions reported in this work (closed symbols) and in \cite{deep2012PLB,deepCompPLB} (open symbols).}
\label{fig:corrections}
\end{figure}

The dependence of the polarization transfer components and their ratios on $Q^2$ (due to the form factors) and other kinematic effects are reduced by dividing the measured values by the expected ones for elastic $\vec ep$ scattering with similar kinematics \cite{deep2012PLB,ceepLet,deepCompPLB,movingProtonPLB}.  
We use the ``moving-proton'' prescription \cite{ArenhovelMoving,movingProtonPLB}, rather than the expressions for elastic scattering off a resting proton \cite{Akh74}.  In this prescription, each measured quasi-elastic event is compared to an elastic event that has the same incident electron
energy ($k_0$), the same magnitude of the four-momentum transfer ($Q^2$), and the same initial proton momentum ($-\vec p_{\rm miss}$)
as the measured quasi-elastic event. Full details on the expressions for the polarization transfer observables  are given in \cite{ArenhovelMoving,movingProtonPLB}.  In Fig.~\ref{fig:corrections}, we compare
the polarization-transfer components calculated for the resting-and the moving-proton kinematics. We show in this figure the ratios $\frac{(P_x)^{^1\!\rm H}_{\rm moving}}{(P_x)^{^1\!\rm H}_{\rm resting}}$, $\frac{(P_z)^{^1\!\rm H}_{\rm moving}}{(P_z)^{^1\!\rm H}_{\rm resting}}$, and $\frac{(P_x/P_z)^{^1\!\rm H}_{\rm moving}}{(P_x/P_z)^{^1\!\rm H}_{\rm resting}}$,
as calculated for the kinematics of the measured $^2$H($\vec e,e'\vec p$) events.  We note that for new kinematics (E) the effect of the moving free proton prescription increases $P_x$ and decreases $P_z$  (unlike the effect in the other kinematics), resulting in a larger effect on $P_x/P_z$.  More details are found in the supplementary.

Following the convention used in \cite{deep2012PLB,ceepLet,deepCompPLB,movingProtonPLB} we use the struck-proton virtuality (``off-shellness''), $\nu$, as our parameter of choice.  The virtuality is defined as \cite{deep2012PLB}:
\begin{equation}
\nu\equiv\Big(m_{A}c-\sqrt{m_{A-1}^{2}c^{2}+p_{\text{miss}}^{2}}\Big)^{2}-p_{\text{miss}}^{2}-m_{p}^{2}c^{2},
\end{equation}
where $m_p$, $m_{A}$ and $m_{A-1}$ are the masses of the proton,  target nucleus (deuteron) and residual nucleus (neutron, assumed to be on-shell) respectively.  The virtuality has been shown to be a more useful parameter than $p_{\rm miss}$ when comparing the double ratios $\frac{(P_x/P_z)^A}{(P_x/P_z)^{^1\!\rm H}}$ in data sets for different nuclei and with different kinematics \cite{deep2012PLB,deepCompPLB,ceepLet}.  The virtuality dependence of the polarization transfer can be different for positive and negative $p_{\rm miss}$ \cite{deep2012PLB,ceepLet,deepCompPLB}. 

\begin{figure}[h!]
\begin{center}
\includegraphics[width=\columnwidth]{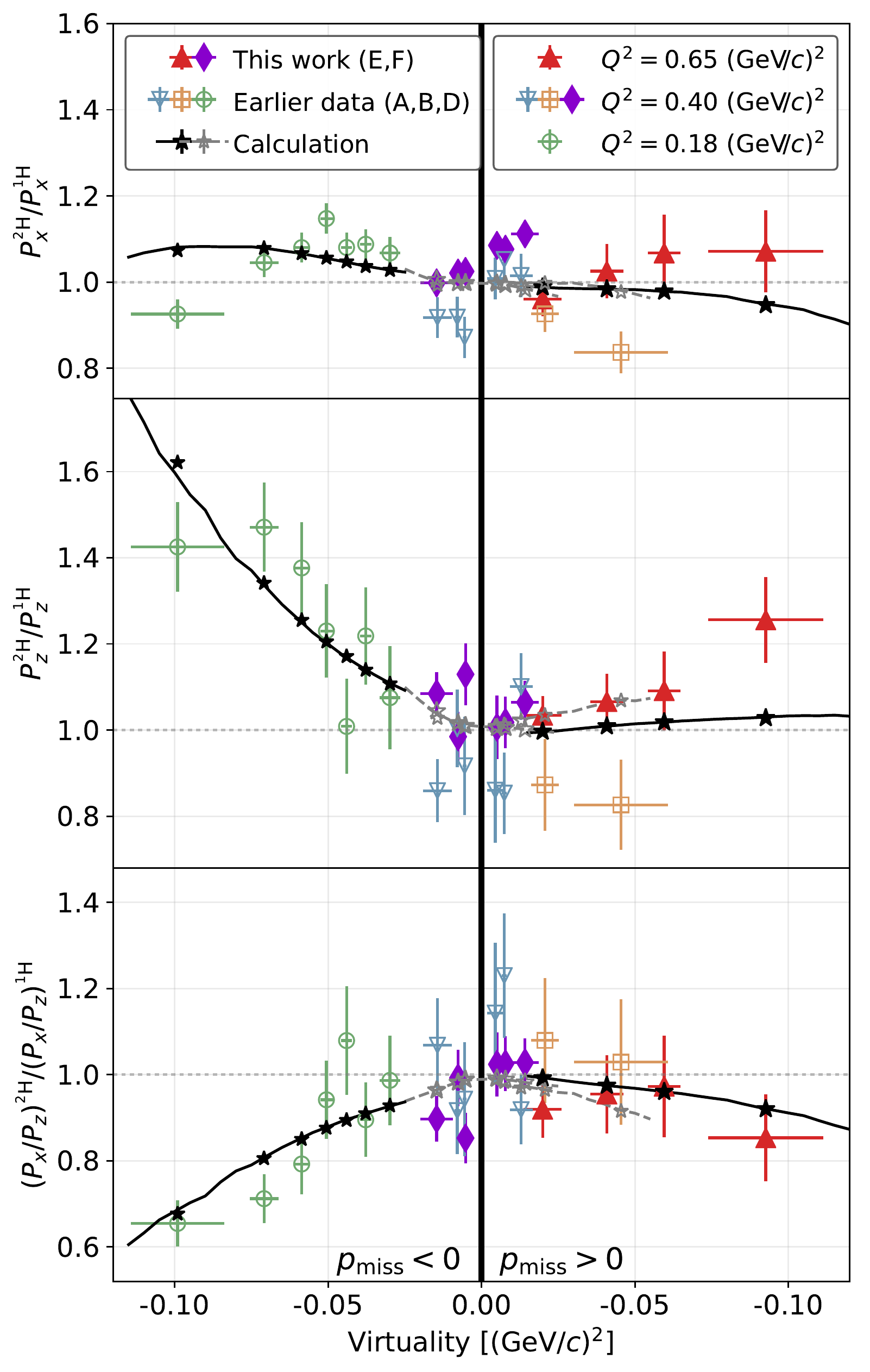}
\caption{
The ratios of the measured $P_x$, $P_z$, and $P_x/P_z$ to those calculated event-by-event for elastic $\vec ep$ scattering using the ``moving proton'' prescription \cite{ArenhovelMoving,movingProtonPLB}.  
Shown on the right (left) side are the data with positive (negative) $p_{\rm miss}$. The new data are
presented with full symbols (red and purple online). The earlier $^2$H data \cite{deep2012PLB,deepCompPLB} are shown as open
symbols. 
The solid black curve is the full calculation (DWIA+MEC+IC(RC)) \cite{Arenhovel} described in Section \ref{sec:calculations} for sets D and E (the other sets are represented by dashed grey curve).  Also shown is this calculation
performed event-by-event and averaged for each bin, presented as full (open) stars for sets D and E (other sets).
}
\label{fig:compDeep}
\end{center}
\end{figure}

We extracted the ratios of the polarization-transfer components for the deuteron to those of a free proton,  $\frac{(P_x)^{^2\!\rm H}}{(P_x)^{^1\!\rm H}}$ and $\frac{(P_z)^{^2\!\rm H}}{(P_z)^{^1\!\rm H}}$, as well as the double ratio $\frac{(P_x/P_z)^{^2\!\rm H}}{(P_x/P_z)^{^1\!\rm H}}$, where the denominator expressions are calculated  event-by-event using free moving-proton kinematics \cite{ArenhovelMoving,movingProtonPLB}.  These ratios are shown as functions of $\nu$ in Fig.~\ref{fig:compDeep}, and compared with those from previous MAMI $^2$H data sets \cite{deep2012PLB,deepCompPLB}.  We find that the differences between the measured polarizations and those of elastic $\vec ep$ scattering in setting E are relatively small compared to those of negative $p_{\rm miss}$ measurement (setting D), where the virtuality is comparably large but the $Q^2$ is smaller.  This suggests that FSI effects are smaller in setting E than in the other settings.  

\begin{figure}[h!]
\begin{center}
\includegraphics[width=\columnwidth]
{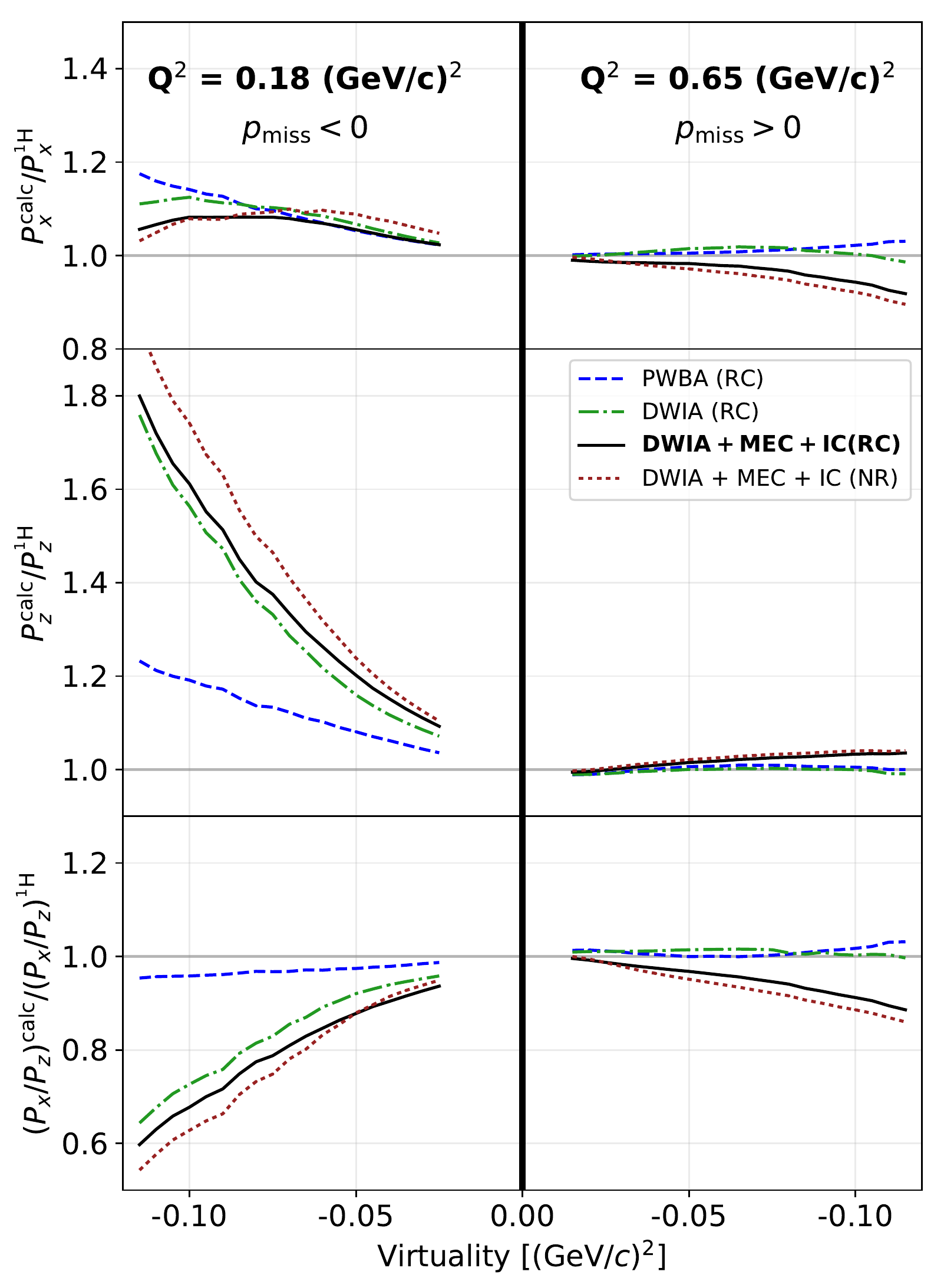}
\caption{The predicted ratios between polarization-transfer observables calculated for the deuteron to those of a free moving proton \cite{ArenhovelMoving,movingProtonPLB}, $\frac{(P_x)^{^2\!\rm H}}{(P_x)^{^1\!\rm H}}$, $\frac{(P_z)^{^2\!\rm H}}{(P_z)^{^1\!\rm H}}$ and $\frac{(P_x/P_z)^{^2\!\rm H}}{(P_x/P_z)^{^1\!\rm H}}$, with various effects included/excluded in the calculations.  On the right side, the kinematics correspond to data set E of this work, while those on the left correspond to those of data set D in Refs.~\cite{deep2012PLB,deepCompPLB}.  The full model is shown as a solid black curve.  See text for details.}
\label{fig:models}
\end{center}
\end{figure}

We compared our measurements to state-of-the-art calculations of the expected polarization transfer for quasi-elastic $\vec ed$ scattering \cite{Arenhovel}. These calculations are based on a realistic potential for the wave functions, i.e. FSI  for the scattering state, and include meson exchange (MEC), isobar (IC) currents, and relativistic contributions (RC) of lowest order. For the nucleon form factors, the parameterizations of Bernauer \textit{et al}. \cite{Bernauer} are used. 
The ratios of the calculated $\vec ed$ polarization-transfer to those of elastic $\vec ep$ scattering are shown as curves\footnote{We use solid black lines for settings D and E in Fig.~\ref{fig:compDeep}, for consistency with Fig.~\ref{fig:models}.  The calculations for the other settings in Fig.~\ref{fig:compDeep} are shown as grey dashed lines.} in Fig.~\ref{fig:compDeep}, along with our measurements.  Also shown are these ratios calculated event-by-event, and then averaged in each bin.  These are shown as filled (open) stars for sets D and E (sets A,B and F). 

We find that the measured $P_x/P_z$ for all the data sets agree very well with the calculations.  The measured components $P_x$ and $P_z$  for both data sets E and F are on average about 3\% above the calculations, which is consistent with the systematic uncertainty on the normalization due to the beam-polarization.  Accounting for this, the measured components for set E agree with the calculation within the errors with $p$-value = 0.11, which is significantly better than in the other kinematic settings\footnote{See the supplemental materials for details.}.

In order to determine which phenomena have the strongest effects on the calculated polarizations, we ran several variations of these calculations, with different effects (FSI, etc.) included or excluded.  The results of these calculations, divided by those of elastic scattering \cite{movingProtonPLB,ArenhovelMoving}, are shown in Fig.~\ref{fig:models}.  The calculations on the right correspond to the kinematics of setting E in this work, with $Q^2 = 0.65$ $({\rm GeV}\!/\!c)^2$ and $p_{\rm miss} > 0$, while those on the left side of the figure correspond to the kinematics of setting D \cite{deep2012PLB,deepCompPLB}, with $Q^2 = 0.18$ $({\rm GeV}\!/\!c)^2$ and $p_{\rm miss} < 0$.
The solid black line is the full calculation in the distorted-wave impulse approximation (DWIA) which includes MEC, IC, and relativistic corrections.  It is identical to that of Fig.~\ref{fig:compDeep}.  The FSI effects, which are included in the DWIA but not in the plane-wave Born approximation (PWBA), strongly affect the $P_z$ component in the $p_{\rm miss}<0$ range, and are greatly reduced in the positive $p_{\rm miss}$ region.  
The combined effects of MEC, IC, and RC, do not exceed 15\%.  
\FloatBarrier

The predicted differences between the \textit{full} quasi-elastic calculation and elastic scattering are very small when the virtuality is near zero, and increase with larger virtuality.  
In general the deviations on the right-hand side, representing the presently explored kinematics (larger $Q^2$, $p_{\rm miss}>0$), are smaller than than those on the left-hand side \cite{deep2012PLB,deepCompPLB}, (low $Q^2$, $p_{\rm miss}<0$).  Moreover, the large increase observed for $P_z$ in the negative $p_{\rm miss}$ range is largely eliminated in the the present kinematics.  This makes calculations in this kinematic range less sensitive to the included effects, and these kinematics more suitable for looking for new phenomena.

In order to estimate limits on possible medium modifications, we reran the calculations on all of the MAMI $^2$H data sets, scaling the form factor ratio $G_E/G_M$ from \cite{Bernauer} by a constant factor $R$.  
For each value of $R$ we compared the calculations to our measurements and computed the associated $p$-value.  From this, we calculated the range in $R$ for which the $p$-value was $\ge$ 5\% (95\% confidence level).  
We accounted for the systematic errors by adding them (multiplied by 1.96 for 95\% confidence) in quadrature with the statistical errors.  
These systematic errors include 2\% for the measurement of $P_x/P_z$ and another 0.5\% for the uncertainty of the $G_E/G_M$ parameterization \cite{Bernauer}.   

\begin{figure}[h!]
\begin{center}
\includegraphics[width=\columnwidth]
{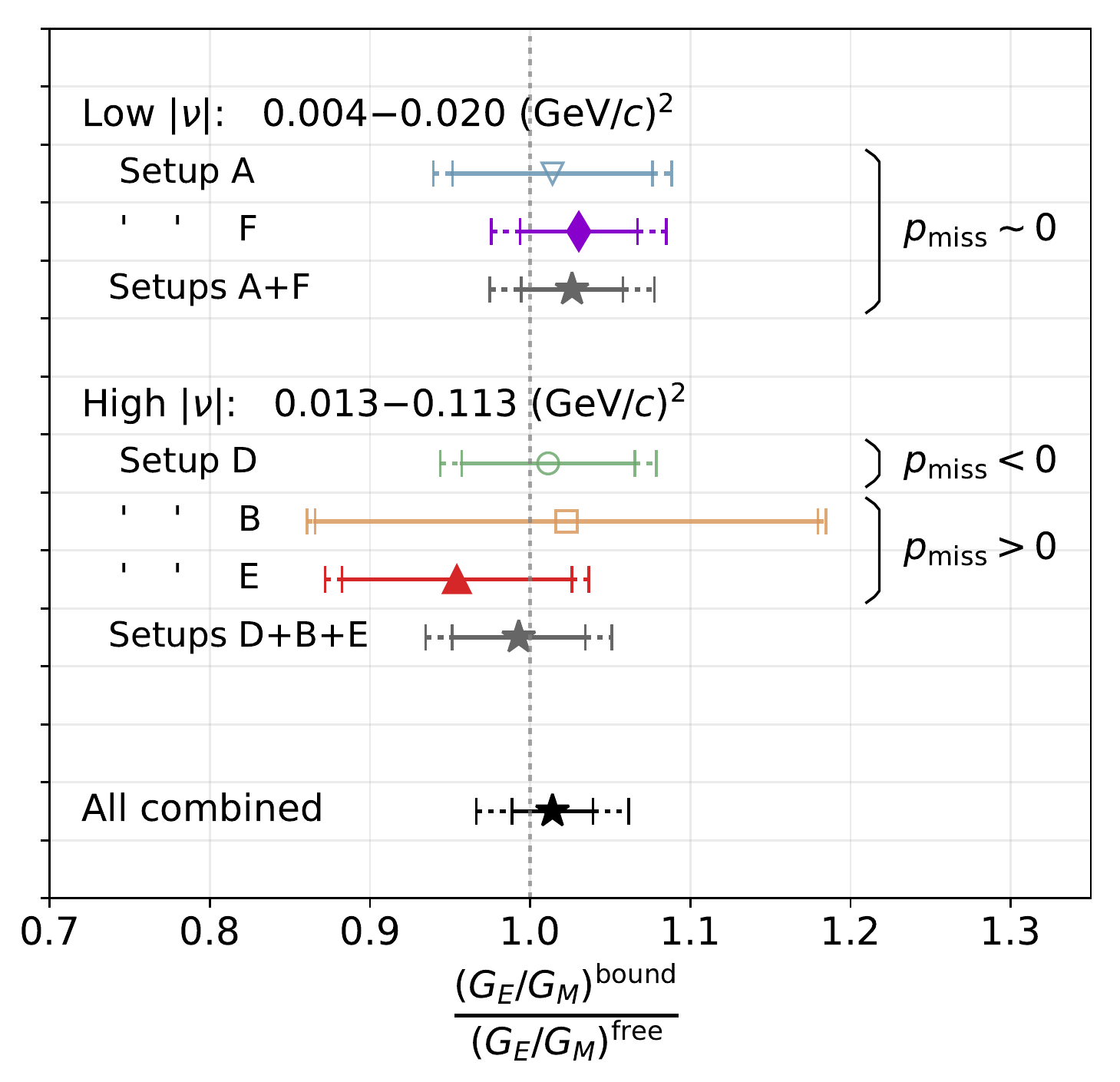}
\caption{95\%-confidence intervals of the ratio $(\frac{G_E}{G_M})^{\rm bound}/(\frac{G_E}{G_M})^{\rm free}$ determined for the present measurements (filled symbols) and those of Refs.~\cite{deep2012PLB,deepCompPLB} (open symbols).  Thick lines denote the limits using statistical errors only, while the dashed lines include both statistical and systematic uncertainties.  data sets with similar ranges in $p_{\rm miss}$ and virtuality are grouped together.  Results from combinations of data sets are shown in black.}
\label{fig:upperLimits}
\end{center}
\end{figure}

The results for all of the MAMI $^2$H data sets (this work and those of \cite{deep2012PLB,deepCompPLB}) are shown in Fig.~\ref{fig:upperLimits}.  We find that each of these data sets is consistent with the hypothesis $R=1$.  
The 95\% confidence interval on $R$ for the combined data set is from 0.966 to 1.062.  This implies that medium-modifications effects on $G_E/G_M$ (if they exist) are constrained to be between $-$3.4\% and 6.2\%.  
This does not exclude medium modifications that keep $G_E/G_M$ constant.

\FloatBarrier
\section{Universality of  the double ratio $\frac{(P_x/P_z)^A}{(P_x/P_z)^{^1\!\rm H}}$}
\label{sec:universality}

\begin{figure}[ht]
\begin{center}
\includegraphics[width=\columnwidth]{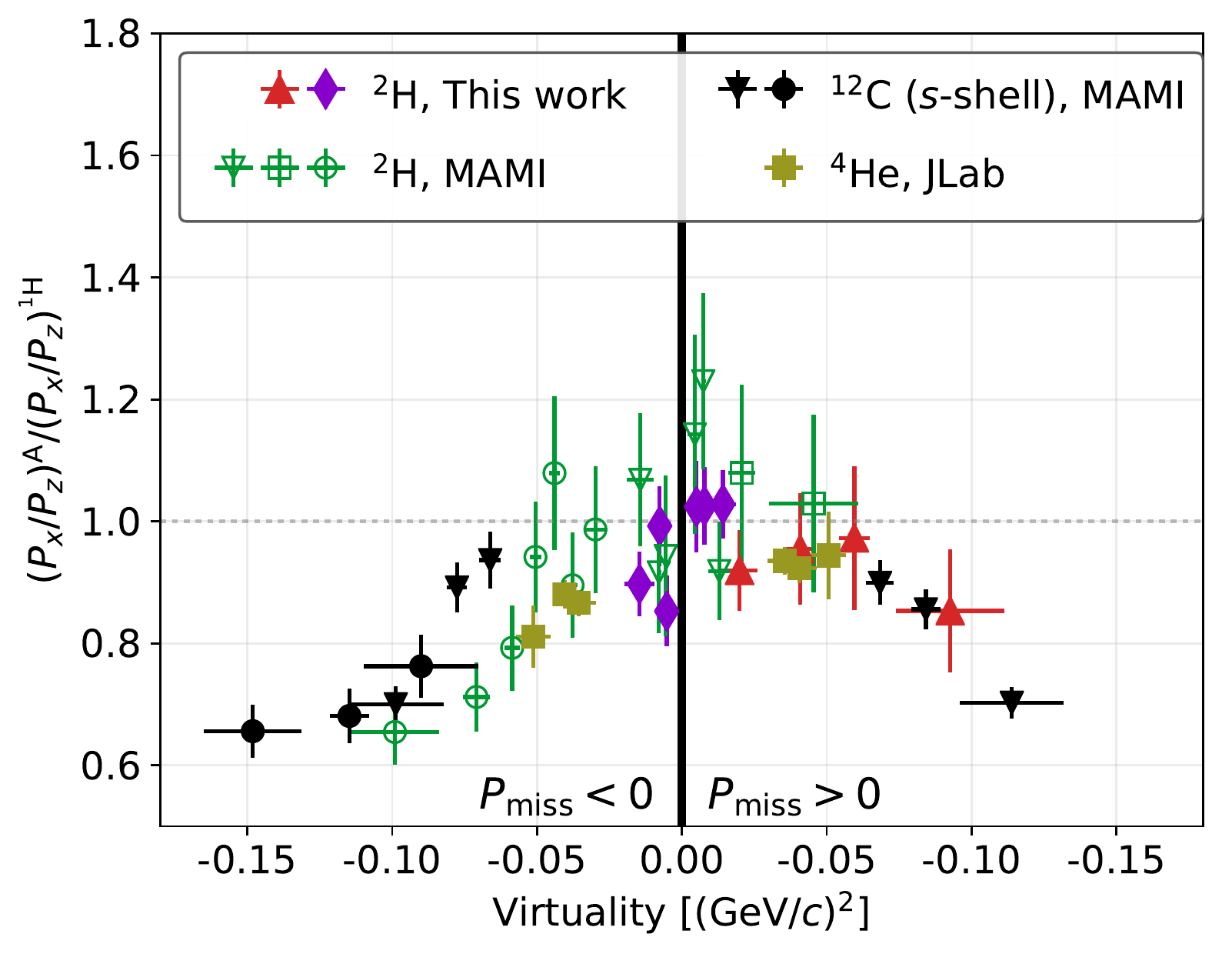}
\caption{
The double-ratios $\frac{(P_x/P_z)^A}{(P_x/P_z)^{^1\!\rm H}}$ measured in this data set (red and purple filled symbols), compared to the earlier $^2$H data sets \cite{deep2012PLB,deepCompPLB} (green, open symbols),  $s$-shell knockout in $^{12}$C data sets \cite{ceepLet} (black filled symbols), and $^4$He data \cite{Paolone} (yellow filled squares).  In each of these ratios, the denominator is calculated using the ``moving-proton'' prescription \cite{ArenhovelMoving,movingProtonPLB}.  The $^4$He data shown was taken at $Q^2$= 0.8 $({\rm GeV}\!/\!c)^2$.}
\label{fig:ceep_deep}
\end{center}
\end{figure}

A good agreement between the double ratios $\frac{(P_x/P_z)^A}{(P_x/P_z)^{^1\!\rm H}}$ in the earlier $^2$H data at $p_{\rm miss}<0$ from Refs.~\cite{deep2012PLB,deepCompPLB} and those measured for $^{12}$C at MAMI in Ref.~\cite{ceepLet} and for $^{4}$He at JLab \cite{Paolone}, was reported in Refs.~\cite{deep2012PLB,ceepLet}. 
To test if the new $^2$H data at $p_{\rm miss}>0$ are likewise compatible with similar measurements in other nuclei at the same virtuality range, we also compare the double-ratios for the new $^2$H data and to those of the previous measurements on $^{4}$He at JLab\footnote{The double ratios presented in \cite{Paolone} have been adapted to use moving proton kinematics \cite{ArenhovelMoving,movingProtonPLB} in their denominator, for a consistent comparison with the MAMI data.  See the supplementary material.} \cite{Paolone}, and $s$-shell knockout\footnote{Polarization transfer in $p_{3/2}$-shell knockout was also measured in \cite{ceepLet}; however, since Fig.~~\ref{fig:ceep_deep} compares the carbon data set with  $^2$H and $^4$He, which only contain $s$-shell protons, only the $s$-shell knockout from carbon is shown in the figure.} on $^{12}$C \cite{ceepLet} at MAMI.  As shown in Fig.~\ref{fig:ceep_deep},  
we find that the double ratio in the new $^2$H data agrees very well with the $^4$He and $^{12}$C in the positive $p_{\rm miss}$ region, just as it was shown for the earlier $^2$H data in Refs.~\cite{ceepLet,movingProtonPLB}  in the negative $p_{\rm miss}$ region.  

The inclusion of setting E in the test of the universality of the double-ratios (Figure \ref{fig:ceep_deep}) reveals the importance of accounting for the proton's initial momentum \cite{movingProtonPLB,ArenhovelMoving} in comparing the measured polarization ratio $P_x/P_z$ to that of elastic scattering.  The moving-proton prescription, introduced in \cite{movingProtonPLB,ArenhovelMoving}, supersedes the earlier ``resting-proton'' prescription, where the bound proton's Fermi motion was not taken into account, and the elastic kinematics 
assumed the proton to initially be at rest.  In \cite{movingProtonPLB}, we found that the effect of using the moving instead of the resting prescription on the double ratio was small for the kinematics available at the time ($\sim$3\% for negative $p_{\rm miss}$, and up to $\sim$10\% for positive $p_{\rm miss}$).  However, we found that this effect is much larger for kinematic setting E, causing the double ratio to increase to 1.5 times larger in the ``resting'' prescription than in the moving prescription, as shown in the supplementary material.  Therefore, the use of the moving (rather than the resting) kinematics is a necessary ingredient in maintaining the universality of the double ratio $\frac{(P_x/P_z)^A}{(P_x/P_z)^{^1\!\rm H}}$.

\FloatBarrier

\section{Conclusions}
We have observed that the differences in the polarization-transfer components between the deuteron and hydrogen are small in the measured region of positive $p_{\rm miss}$ and large $Q^2$, compared to previous measurements at negative  $p_{\rm miss}$ and small $Q^2$, as predicted by the calculations.  This implies that the effects of FSI are small in this region, making it a good region for searching for medium modifications.

The measured polarization-components $P_x$ and $P_z$ for the new $^2$H data, as well as their ratio, $P_x/P_z$, are consistent with the calculations within the errors.  This tests the validity of the calculations on the positive missing-momentum region, and shows no evidence of medium modifications on the FF ratio $G_E/G_M$.  Further, we have established constraints on possible modifications to this ratio on the order of a few percent.

We also find that, when dividing our measured values by the polarization-transfer for a free moving proton \cite{ArenhovelMoving,movingProtonPLB}, the double ratios $\frac{(P_x/P_z)^A}{(P_x/P_z)^{^1\!\rm H}}$ from the $^2$H measurements in the new kinematic region are in agreement with measurements for $^{4}$He and $^{12}$C, just as they were found to be in agreement in previous $^2$H measurements.  This supports the universal behavior of the double ratio.

\section{Acknowledgements}
We would like to thank the Mainz Microtron operators and technical staff for the excellent operation of the accelerator.  
This work is
supported by the Israel Science Foundation (grant 390/15) of the Israel
Academy of Arts and Sciences, 
by the PAZY Foundation (grant 294/18),
by the Deutsche Forschungsgemeinschaft (Collaborative Research
Center 1044), by the Slovenian Research Agency (research core funding
No.~P1\textendash 0102), by the U.S. National Science Foundation
(PHY-1505615), and by the Croatian Science Foundation Project No.~8570.

\section{Supplementary materials}
Supplementary materials may be found online at \url{https://doi.org/10.1016/j.physletb.2019.07.002}.

\section*{References}
\bibliographystyle{elsarticle-num}
\addcontentsline{toc}{section}{\refname}\small{\bibliography{comp}}

\begin{thebibliography}{10}
\expandafter\ifx\csname url\endcsname\relax
  \def\url#1{\texttt{#1}}\fi
\expandafter\ifx\csname urlprefix\endcsname\relax\def\urlprefix{URL }\fi
\expandafter\ifx\csname href\endcsname\relax
  \def\href#1#2{#2} \def\path#1{#1}\fi

\bibitem{Jones:1999rz}
M.~K. Jones, et~al., { $G_{Ep} / G_{Mp}$ ratio by polarization transfer in
  polarized $\vec ep\!\rightarrow \!e \vec p$}, Phys. Rev. Lett. 84 (2000)
  1398--1402.
\newblock \href {http://arxiv.org/abs/nucl-ex/9910005}
  {\path{arXiv:nucl-ex/9910005}}, \href
  {http://dx.doi.org/10.1103/PhysRevLett.84.1398}
  {\path{doi:10.1103/PhysRevLett.84.1398}}.

\bibitem{Gayou:2001qd}
O.~Gayou, et~al., {Measurement of $G_{Ep} / G_{Mp}$ in $\vec ep\rightarrow e
  \vec p$ to $Q^2$ = 5.6 GeV$^2$}, Phys. Rev. Lett. 88 (2002) 092301.
\newblock \href {http://arxiv.org/abs/nucl-ex/0111010}
  {\path{arXiv:nucl-ex/0111010}}, \href
  {http://dx.doi.org/10.1103/PhysRevLett.88.092301}
  {\path{doi:10.1103/PhysRevLett.88.092301}}.

\bibitem{Punjabi:2005wq}
V.~Punjabi, et~al., {Proton elastic form-factor ratios to Q$^2$ = 3.5-GeV$^2$
  by polarization transfer}, Phys. Rev. C 71 (2005) 055202, [Erratum: Phys.
  Rev.C71,069902(2005)].
\newblock \href {http://arxiv.org/abs/nucl-ex/0501018}
  {\path{arXiv:nucl-ex/0501018}}, \href
  {http://dx.doi.org/10.1103/PhysRevC.71.055202, 10.1103/PhysRevC.71.069902}
  {\path{doi:10.1103/PhysRevC.71.055202, 10.1103/PhysRevC.71.069902}}.

\bibitem{Milbrath:1997de}
B.~D. Milbrath, et~al., {A Comparison of polarization observables in electron
  scattering from the proton and deuteron}, Phys. Rev. Lett. 80 (1998)
  452--455, [Erratum: Phys. Rev. Lett. 82, 2221 (1999)].
\newblock \href {http://arxiv.org/abs/nucl-ex/9712006}
  {\path{arXiv:nucl-ex/9712006}}, \href
  {http://dx.doi.org/10.1103/PhysRevLett.80.452, 10.1103/PhysRevLett.82.2221}
  {\path{doi:10.1103/PhysRevLett.80.452, 10.1103/PhysRevLett.82.2221}}.

\bibitem{Barkhuff:1999xc}
D.~H. Barkhuff, et~al., {Measurement of recoil proton polarizations in the
  electrodisintegration of deuterium by polarized electrons}, Phys. Lett. B 470
  (1999) 39--44.
\newblock \href {http://dx.doi.org/10.1016/S0370-2693(99)01294-0}
  {\path{doi:10.1016/S0370-2693(99)01294-0}}.

\bibitem{Pospischil:2001pp}
T.~Pospischil, et~al., {Measurement of $G_{Ep}/G_{Mp}$ via polarization
  transfer at $Q^2$ = 0.4 (GeV$/c)^2$}, Eur. Phys. J. A 12 (2001) 125--127.
\newblock \href {http://dx.doi.org/10.1007/s100500170046}
  {\path{doi:10.1007/s100500170046}}.

\bibitem{PhysRevC.64.038202}
O.~Gayou, K.~Wijesooriya, et~al., Measurements of the elastic electromagnetic
  form factor ratio ${\ensuremath{\mu}}_{p}{G}_{\mathrm{ep}}{/G}_{\mathrm{mp}}$
  via polarization transfer, Phys. Rev. C 64 (2001) 038202.
\newblock \href {http://dx.doi.org/10.1103/PhysRevC.64.038202}
  {\path{doi:10.1103/PhysRevC.64.038202}}.

\bibitem{MACLACHLAN2006261}
G.~MacLachlan, et~al., The ratio of proton electromagnetic form factors via
  recoil polarimetry at $q^2$=1.13 $({\rm gev}/c)^2$, Nuclear Physics A 764
  (2006) 261 -- 273.
\newblock \href {http://dx.doi.org/10.1016/j.nuclphysa.2005.09.012}
  {\path{doi:10.1016/j.nuclphysa.2005.09.012}}.

\bibitem{PhysRevC.74.035201}
M.~K. Jones, et~al., Proton ${G}_{E}/{G}_{M}$ from beam-target asymmetry, Phys.
  Rev. C 74 (2006) 035201.
\newblock \href {http://dx.doi.org/10.1103/PhysRevC.74.035201}
  {\path{doi:10.1103/PhysRevC.74.035201}}.

\bibitem{Akh74}
A.~I. Akhiezer, M.~Rekalo,
  \href{http://refhub.elsevier.com/S0370-2693(17)30052-7/bib416B683734s1}{Polarization
  effects in the scattering of leptons by hadrons}, Sov. J. Part. Nucl. 4
  (1974) 277, [Fiz. Elem. Chast. Atom. Yadra 4, (1973) 662].

\bibitem{Perdrisat}
C.~F. Perdrisat, et~al., {Nucleon electromagnetic form factors}, Prog. Part.
  Nucl. Phys. 59 (2007) 694--764.
\newblock \href {http://dx.doi.org/10.1016/j.ppnp.2007.05.001}
  {\path{doi:10.1016/j.ppnp.2007.05.001}}.

\bibitem{Arenhovel}
H.~Arenh{\"o}vel, W.~Leidemann, E.~L. Tomusiak, {General survey of polarization
  observables in deuteron electrodisintegration}, Eur.\ Phys.\ J. A 23 (2005)
  147--190.
\newblock \href {http://dx.doi.org/10.1140/epja/i2004-10061-5}
  {\path{doi:10.1140/epja/i2004-10061-5}}.

\bibitem{deep2012PLB}
I.~Yaron, D.~Izraeli, et~al., Polarization-transfer measurement to a
  large-virtuality bound proton in the deuteron, Phys.\ Lett.\ B 769 (2017)
  21--24.
\newblock \href {http://dx.doi.org/10.1016/j.physletb.2017.01.034}
  {\path{doi:10.1016/j.physletb.2017.01.034}}.

\bibitem{deepCompPLB}
D.~Izraeli, I.~Yaron, et~al., {Components of polarization-transfer to a bound
  proton in a deuteron measured by quasi-elastic electron scattering}, Phys.
  Lett. B 781 (2018) 107--111.
\newblock \href {http://arxiv.org/abs/1801.01306} {\path{arXiv:1801.01306}},
  \href {http://dx.doi.org/10.1016/j.physletb.2018.03.063}
  {\path{doi:10.1016/j.physletb.2018.03.063}}.

\bibitem{ceepLet}
D.~Izraeli, T.~Brecelj, et~al., {Measurement of polarization-transfer to bound
  protons in carbon and its virtuality dependence}, Phys. Lett. B 781 (2018)
  95--98.
\newblock \href {http://arxiv.org/abs/1711.09680} {\path{arXiv:1711.09680}},
  \href {http://dx.doi.org/10.1016/j.physletb.2018.03.027}
  {\path{doi:10.1016/j.physletb.2018.03.027}}.

\bibitem{jlabDeep}
B.~Hu, et~al., {Polarization transfer in the $^2$H($\vec e, e' \vec p$)n
  reaction up to $Q^2$ = 1.61 (GeV/c)$^2$}, Phys.\ Rev. C 73 (2006) 064004.
\newblock \href {http://dx.doi.org/10.1103/PhysRevC.73.064004}
  {\path{doi:10.1103/PhysRevC.73.064004}}.

\bibitem{Strauch}
S.~Strauch, et~al., {Polarization transfer in the ${^4}$\textsc{H}e$(\vec
  e,e'\vec p)$${^3}$\textsc{H} reaction up to Q${^2}$ = 2.6~(GeV/$c$)${^2}$},
  Phys. Rev. Lett. 91 (2003) 052301.
\newblock \href {http://dx.doi.org/10.1103/PhysRevLett.91.052301}
  {\path{doi:10.1103/PhysRevLett.91.052301}}.

\bibitem{Paolone}
M.~Paolone, S.~P. Malace, S.~Strauch, et~al., Polarization transfer in the
  $^{4}\mathrm{He}(\vec e,e'\vec p)^{3}\mathbf{H}$ reaction at ${Q}^{2}=0.8$
  and $1.3\text{ }\text{ }(\mathrm{GeV}/c{)}^{2}$, Phys. Rev. Lett. 105 (2010)
  072001.
\newblock \href {http://dx.doi.org/10.1103/PhysRevLett.105.072001}
  {\path{doi:10.1103/PhysRevLett.105.072001}}.

\bibitem{Malov_O16}
S.~Malov, et~al., Polarization transfer in the $^{16}\mathrm{O}(\vec e,e'\vec
  p)^{15}\mathrm{N}$ reaction, Phys.\ Rev.\ C 62 (2000) 057302.
\newblock \href {http://dx.doi.org/10.1103/PhysRevC.62.057302}
  {\path{doi:10.1103/PhysRevC.62.057302}}.

\bibitem{a1aparatus}
K.~Blomqvist, et~al., The three-spectrometer facility at {MAMI}, Nucl.\
  Instrum.\ and Meth.\ A 403~(2--3) (1998) 263 -- 301.
\newblock \href {http://dx.doi.org/10.1016/S0168-9002(97)01133-9}
  {\path{doi:10.1016/S0168-9002(97)01133-9}}.

\bibitem{Wagner}
B.~Wagner, H.~G. Andresen, K.~H. Steffens, W.~Hartmann, W.~Heil, E.~Reichert,
  {A Moller polarimeter for CW and pulsed intermediate-energy electron beams},
  Nucl. Instrum. Meth. A294 (1990) 541--548.
\newblock \href {http://dx.doi.org/10.1016/0168-9002(90)90296-I}
  {\path{doi:10.1016/0168-9002(90)90296-I}}.

\bibitem{Bartsch}
P.~Bartsch,
  \href{http://wwwa1.kph.uni-mainz.de/A1/publications/doctor/bartsch.pdf}{{Aufbau
  eines M\o ller-polarimeters f\"ur die drei-spektrometer-anlage und messung
  der helizit\"atsasymmetrie in der reaktion $p(e, e' p)\pi_0$ im bereich der
  $\Delta$-resonanz}}, Ph.D. thesis, {Institut f\"ur Kernphysik der
  Universit\"at Mainz} (2001).
\newline\urlprefix\url{http://wwwa1.kph.uni-mainz.de/A1/publications/doctor/bartsch.pdf}

\bibitem{Tioukine}
V.~Tioukine, K.~Aulenbacher, E.~Riehn, et~al., {A Mott polarimeter operating at
  MeV electron beam energies}, Rev. Sci. Instrum. 82~(3) (2011) 033303.
\newblock \href {http://dx.doi.org/10.1063/1.3556593}
  {\path{doi:10.1063/1.3556593}}.

\bibitem{Pospischil:2000pu}
T.~Pospischil, et~al., The focal plane proton-polarimeter for the
  3-spectrometer setup at {MAMI}, Nucl.\ Instrum.\ Methods.\ Phys.\ Res.,
  Sect.\ A 483~(3) (2002) 713 -- 725.
\newblock \href {http://dx.doi.org/10.1016/S0168-9002(01)01955-6}
  {\path{doi:10.1016/S0168-9002(01)01955-6}}.

\bibitem{AprileGiboni:1984pb}
E.~Aprile-Giboni, R.~Hausammann, E.~Heer, R.~Hess, C.~Lechanoine-Leluc, W.~Leo,
  S.~Morenzoni, Y.~Onel, D.~Rapin, {Proton-carbon effective analyzing power
  between 95 and 570 MeV}, Nucl. Instrum. Meth. 215 (1983) 147--157.
\newblock \href {http://dx.doi.org/10.1016/0167-5087(83)91302-9}
  {\path{doi:10.1016/0167-5087(83)91302-9}}.

\bibitem{Mcnaughton:1986ks}
M.~W. McNaughton, et~al., {The p-C analyzing power between 100 and 750 MeV},
  Nucl. Instrum. Meth. A 241 (1985) 435--440.
\newblock \href {http://dx.doi.org/10.1016/0168-9002(85)90595-9}
  {\path{doi:10.1016/0168-9002(85)90595-9}}.

\bibitem{movingProtonPLB}
S.~Paul, T.~Brecelj, H.~Arenh{\"o}vel, et~al., {The influence of Fermi motion
  on the comparison of the polarization transfer to a proton in elastic $\vec
  ep$ and quasi-elastic $\vec eA$ scattering}, Phys. Lett. B 792 (2019)
  445--449.
\newblock \href {http://arxiv.org/abs/1901.10958} {\path{arXiv:1901.10958}},
  \href {http://dx.doi.org/10.1016/j.physletb.2019.04.004}
  {\path{doi:10.1016/j.physletb.2019.04.004}}.

\bibitem{ArenhovelMoving}
H.~Arenh{\"o}vel, {Polarization observables for elastic electron scattering off
  a moving nucleon}, Phys. Rev. C 99 (2019) 055502.
\newblock \href {http://arxiv.org/abs/1904.04515} {\path{arXiv:1904.04515}},
  \href {http://dx.doi.org/10.1103/PhysRevC.99.055502}
  {\path{doi:10.1103/PhysRevC.99.055502}}.

\bibitem{Bernauer}
J.~C. Bernauer, et~al., {Electric and magnetic form factors of the proton},
  Phys.\ Rev. C 90~(1) (2014) 015206.
\newblock \href {http://dx.doi.org/10.1103/PhysRevC.90.015206}
  {\path{doi:10.1103/PhysRevC.90.015206}}.

\end{thebibliography}
\end{document}